# Visualization of Electron Nematicity and Unidirectional Antiferroic Fluctuations at High Temperatures in NaFeAs

E. P. Rosenthal[1], E. F. Andrade[1], C. J. Arguello[1], R. M. Fernandes[2], L. Y. Xing[3], X. C. Wang[3], C. Q. Jin[3], A. J. Millis[1], A. N. Pasupathy[1]

[1]Department of Physics, Columbia University, New York NY 10027, USA

[2]School of Physics and Astronomy, University of Minnesota, Minneapolis, MN 55455, USA

[3]Beijing National Laboratory for Condensed Matter Physics, Institute of Physics, Chinese Academy of Sciences, Beijing 100190, China

**Abstract**:

**The driving forces behind electronic nematicity in the iron pnictides remain hotly debated. We use atomic-resolution variable-temperature scanning tunneling spectroscopy to provide the first direct visual evidence that local electronic nematicity and unidirectional antiferroic (stripe) fluctuations persist to temperatures almost twice the nominal structural ordering temperature in the parent pnictide NaFeAs. Low-temperature spectroscopic imaging of nematically-ordered NaFeAs shows anisotropic electronic features that are not observed for isostructural, non-nematic LiFeAs. The local electronic features are shown to arise from scattering interference around crystalline defects in NaFeAs, and their spatial anisotropy is a direct consequence of the structural and stripe-magnetic order present at low temperature. We show that the anisotropic features persist up to high temperatures in the nominally tetragonal phase of the crystal. The spatial distribution and energy dependence of the anisotropy at high temperatures is explained by the persistence of large amplitude, short-range, unidirectional, antiferroic (stripe) fluctuations, indicating that strong density wave fluctuations exist and couple to near-Fermi surface electrons even far from the structural and density wave phase boundaries.**

**Main Text**:

The nature of the normal state from which superconductivity emerges in unconventional superconductors remains a mystery. It is suspected that electronic interactions present in the normal state play a key role in the formation of the superconducting state[1]. In both the cuprates and the pnictide phase diagrams, magnetically ordered states exist in proximity to the superconducting state, and the pnictides additionally exhibit orbital ordering [2,3]. A crucial additional feature of the pnictides is the appearance of a "nematic'' phase in which the tetragonal rotational symmetry of the ideal pnictide lattice is spontaneously broken below a temperature $T_S$. Recent bulk transport and scattering measurements have suggested that the nematic phase is driven by electronic, rather than lattice, degrees of freedom [4-8] and is observed in all electronic channels – charge [9,10], orbital [4], and spin [7,11]. Spin order and spin fluctuations [12-16] (which couple quadratically to nematicity) as well as orbital order [17,18] and orbital fluctuations [19] (which can couple linearly) have been invoked to explain the nematicity. However, the dominant interaction responsible for the nematic ordering and fluctuations remains unknown and identifying it is a key experimental goal. In this paper, we use variable temperature scanning tunneling spectroscopy to provide new insights into this issue by showing that our spectroscopic signals reveal that nematicity occurs in conjuction with strong antiferroic fluctuations and that both phenomena persist up to temperatures much greater than the temperatures at which long-range order is established.

The arsenide superconductors consist of one or more iron-arsenide layers with the iron atoms in a tetrahedral bonding environment at high temperature [20]. Below a temperature $T_S$, a structural transition occurs to an orthorhombic phase in which both the *a* and *b* directions of the iron-arsenide plane [21] and the occupation of the $3d_{xz}$ and $3d_{yz}$ iron orbitals [4] become inequivalent. In mathematical terms, the symmetry is lowered from $C_4$ (symmetry of the square) to $C_2$ (symmetry only under 180° rotation). At a temperature $T_{SDW} \leq T_S$, magnetic order sets in with a wave vector $\mathbf{Q}_a = (\frac{\pi}{a_0}, 0)$ where $a_0$ is the longer iron–iron distance in the orthorhombic phase [22]. The structural (nematic) and magnetic (stripe) transitions appear to be intimately linked, since $T_{SDW}$ and $T_S$ track each other across the phase diagram, even inside the superconducting dome [23,24]. Recent transport experiments showed that even a small strain imposed on a detwinned sample can induce significant anisotropies in the tetragonal phase, indicating that the susceptibility to nematic order is very large even at temperatures well above the ordering temperature[8]. However, electrical transport measurements average over the entire sample and are related in a complicated way to the specific electronic properties. Scanning tunneling microscopy (STM) studies have successfully visualized anisotropic electronic states in the low temperature orthorhombic state of pnictides[25,26,27]. Because they are performed at low temperatures, though, these measurements are sensitive only to the coupling between nematicity and the saturated order parameter values and do not provide information about fluctuations.

In this paper we use variable temperature scanning tunneling spectroscopy (STS) with atomic scale spatial sensitivity and millivolt energy resolution to visualize the electronic wavefunctions of NaFeAs at temperatures both below and above the magnetic and structural phase boundaries and thus gain new insights into the physics of the ordered electronic nematic state and the high temperature state that gives rise to it. The key to our results is the observation that structural defects give rise to local modulations of the electronic density of states (termed ``quasiparticle interference"—QPI--), which may be detected in STS experiments and analyzed to obtain information about the underlying electronic state[28-30]. We present a comparison of the density of states modulations produced by impurities in the low temperature nematic/magnetic phase of the prototypical pnictide material NaFeAs to those produced in non-nematic LiFeAs; this comparison allows us to determine the effects of nematicity in the STS spectra. We find that the characteristic anisotropic features of the STS spectra reflect both the nematic order and the Fermi surface rearrangement arising from the unidirectional magnetic (stripe) order which coexists with the structural order. By increasing temperature, we find that some key features of the NaFeAs STS data persist to temperatures up to 90 K, more than 1.5 times the structural phase transition temperature of 56 K and more than twice the stripe ordering temperature of 40 K[31,32]. Our theoretical analysis shows that the main features of the STS data at high temperatures are reproduced by a model of electrons in the presence of strong stripe fluctuations.

The 111 family of iron arsenides ([AFeAs] with A an alkali metal) has been shown to be suitable for STM experiments because cleaving produces nonpolar surfaces [29,30,33-35]. We perform our experiments on single crystals of NaFeAs and LiFeAs which have been cleaved in UHV at temperatures between 20 and 100 K. NaFeAs has a structural transition at $T_S$=54 K and a magnetic transition at $T_{SDW}$=39 K. The unusually large temperature separation between these transitions allows us to sensitively probe the electronic structure in all three phases of the material. LiFeAs is chosen as a reference material since it has neither a structural nor a magnetic transition down to low temperature [36-38]. (LiFeAs does have a low-T superconducting state, but our measurements are taken above this transition temperature). An STM topographic image typical of either of these materials (in this case LiFeAs) is shown in Fig. 1a. The square lattice of alkali atoms is observed which is consistent with previous measurements for both materials [29,30,33-35]. The iron atoms (shown schematically in the lower left portion of the figure) lie below the alkali atoms, with in-plane position approximately half-way between pairs of Li atoms. Differences in the position of the out of plane As atoms (not shown) make the two Fe sites inequivalent (indicated by filled and unfilled circles). Large area topographs in both materials reveal the presence of several types of crystalline defects which is typical of all samples studied [33-35]. The STM topographic features produced by some of these defects (such as the alkali vacancy, black arrow, Fig. 1a) do not break the local $C_4$ symmetry, while other defects (such as the iron site defects [35] which are visualized as the bright, yellow, "dumbbell"-shaped features in Fig. 1a) do reduce the local symmetry to $C_2$ in their immediate vicinity. These iron defects produce topographic features whose symmetry axes are along the Alkali-Alkali (A-A) nearest neighbor directions [35] and whose orientation is determined by the iron position underneath the midpoint of the A-A bond.

As expected, all defects modify the local density of states (LDOS) in their immediate vicinity. However, the nature of the changes in the LDOS is very different in LiFeAs and NaFeAs. We illustrate this by showing STM spectroscopic images at an energy of E=10 meV and E=14 meV around two inequivalent iron-site defects in LiFeAs (Fig. 1b) and NaFeAs (Fig. 1c) respectively, where we can see C2 symmetric features in both cases. Fig. 1b, which has been obtained on LiFeAs at 39 K, contains two inequivalent $C_2$ symmetric features that arise from two inequivalent iron defects (see Supplementary Fig. S1 for corresponding topographic image). The spectroscopic signatures of the defects have roughly the same size (about 10 Å) and point in the same directions (Li-Li) as the corresponding topographic features, whose sizes and orientations are indicated by red lines. We find that in LiFeAs there are an equal number of iron defects on average in each of the two inequivalent positions, resulting in an equal number of features oriented along either Li-Li axis. Thus, while the electronic structure of LiFeAs is not $C_4$ symmetric at the atomic scale due to iron site defects, averages over larger length scales of the electronic structure cancel out these differences from the local defect states and become $C_4$ symmetric. Conversely, Figure 1c shows a spectroscopic image of NaFeAs measured at 26 K, deep within the SDW phase. As in Fig. 1b, we identify the topographic size and orientation of two inequivalent iron defects with red lines. Unlike in LiFeAs, the spectroscopic signatures of the defects are both pointed along the same direction, 45° with respect to Na-Na axes, corresponding to the Fe-Fe nearest neighbor direction. These spectroscopic features also have a much greater spatial extent (typically 50 Å) than the topographic signatures of the defects. Consequently, $C_4$ symmetry is broken over large scales compared to the defect separation. The spatial extent and anisotropy of the features obtained on NaFeAs additionally have a pronounced energy dependence as seen in Fig. 1d-i, while the energy dependence in the case of LiFeAs is much weaker (see Supplementary Fig. S1).

A second key difference between LiFeAs and NaFeAs is the relationship between the nature of the defect and the changes in the LDOS in spectroscopic images. In the case of LiFeAs, different defects produce different patterns, and in particular the defects that do not break C4 symmetry in topography do not do so in spectroscopy either (see Supplementary Information). On the other hand, in NaFeAs, the spectroscopic features seen in large area maps are completely independent of the nature of the defect as shown by the spectroscopic survey over a much larger area in Fig. 1j (see also Supplementary Fig. S3). Figure 1j, which contains many types of defects that all produce the same spectroscopic feature, clearly depicts the presence of two domains corresponding to the two inequivalent ways of breaking $C_4$ symmetry. The typical size of domains observed at low temperature is 1-2 µm in our samples, which is consistent with the size of twin domains in pnictides [39]. All of these facts indicate that these unidirectional spectroscopic features in NaFeAs seen in the SDW phase arise from the anisotropic electronic structure of the material.

Given that the LDOS patterns seen in NaFeAs are independent of the nature of the defect and have strong energy dependence, a natural explanation is quasiparticle interference (QPI). In QPI, defects serve as scattering centers and produce interference patterns that change in contrast and wavelength as a function of energy. Accordingly, we examine the two-dimensional Fast Fourier transform (FFT) of conductance images to understand the momenta of quasiparticles involved in the scattering processes [26]. To achieve high Fourier space resolution, we obtained spectroscopic images over a large, 110 nm square field of view. These real space images are shown for select energies in Fig. 2a-d, which demonstrate that the unidirectional features persist over a range of energies around the Fermi energy. The FFTs of the real space images are shown in Fig. 2e-h (FFT procedure described in Supplementary Information). The rich structure seen in the FFTs is strongly energy dependent, as expected from QPI, and we do not see evidence of the energy-independent features in NaFeAs that have been observed in previous STM measurements on $Ca(Fe_{1-x}Co_x)_2As2$ [26,40].

Having seen unidirectional structure in the Scanning Tunneling Spectroscopy (STS) images at low temperature, we ask how the STS images evolve as one raises the temperature beyond $T_{SDW}$ (39 K) and $T_S$ (54 K). We have performed STS measurements on a series of samples at many temperatures between 20 K and 85 K. Multiple samples have been measured at each temperature over a period of several months to rule out tip or sample artifacts, with the tip calibrated on single crystal gold before each sample measurement. Figure 3 plots the raw STS data through both phase transitions at temperatures of 26 K, 38 K, 46 K, 54 K, 61 K, and 75 K along with their respective FFT images. All images are plotted at an energy of E=10 meV and are obtained under similar junction conditions. We can clearly see that the unidirectional features which are seen below $T_{SDW}$=39 K persist to at temperatures far above the structural transition $T_S$=54 K, showing $C_4$ symmetry breaking by low energy electronic states over the STM field of view. The $C_4$ symmetry breaking does get weaker with increasing temperature and is not experimentally measurable at temperatures above ~ 90 K.

A potential cause for concern in interpreting STM data on any bulk crystal is the relationship between bulk and surface electronic structure. We have two pieces of evidence that data obtained from surface probes such as STS are representative of bulk behavior. First, a clear SDW gap is observed in the spectrum (see Ref. [33] and also the discussion below) that disappears near the bulk SDW transition of 39 K. Second, angle-resolved photoemission spectroscopy (ARPES, also a surface probe which averages over a wide area) data from similar crystals [41,42] indicates that the temperature dependent band structure of NaFeAs at the surface is truly representative of the bulk crystal properties. In particular, the orthorhombic to tetragonal transition is clearly observed in ARPES from the temperature-dependence of the bands.

We now consider the implications of our results for the physics of the pnictide materials. The first point, independent of any theoretical analysis, is the observed persistence of local $C_4$ symmetry breaking in the local electronic structure to temperatures ~90K, well above the measured structural phase transition temperature $T_S$= 54K. Our results are consistent with electrical resistivity measurements on detwinned crystals subject to a small externally imposed symmetry breaking strain that report a temperature dependent anisotropy setting in below ~90K which was interpreted as arising from a rapidly growing nematic susceptibility[8,43]. As we now show, the STM data reveals important additional physics beyond the $C_4$ symmetry breaking, namely that strong, short-ranged, antiferroic fluctuations play a key role in the electronic structure even at temperatures up to 90 K, well into the paramagnetic nominally symmetry-unbroken phase.

The key features in the FFT images that we seek to understand are shown in Fig. 4a-c, taken at the Fermi energy at temperatures of 26 K, 46 K and 61 K respectively. At all three temperatures there is a strong signal along rods centered at $\mathbf{Q_x} = 0$ and $\mathbf{Q_x} = \mathbf{q_D} \equiv \pm 0.15 \dfrac{\pi}{a_0}$. The breaking of rotational symmetry is evident in the alignment of the rods (in this image, parallel to the y direction). It will be shown that the structures at $\mathbf{Q_x} = \pm \mathbf{q_D}$ are accounted for by antiferroic fluctuations.

To analyze the FFT images more quantitatively, we turn to the theory of QPI, which states [44] that the observed spatial modulations in dI/dV at energy ω are directly proportional to impurity-induced modulations of the local electronic density of states $\delta N(R,\omega)$. The Fourier transform may be computed from the electron Green's function $G(\mathbf{k},\omega)$, which describes propagation of an electron of energy ω and crystal momentum **k** in a defect-free material, and the T-matrix $T_{\mathbf{r}_{imp}}(\mathbf{k},\mathbf{k}+\mathbf{Q};\omega)$, which describes the total scattering amplitude of an electron from crystal momentum **k** to crystal momentum $\mathbf{k}+\mathbf{Q}$ by a defect at position $\mathbf{r}_{imp}$

$$\delta N(\mathbf{Q},\omega) = \mathrm{Im}\int d\mathbf{k} G(\mathbf{k},\omega) T_{\mathbf{r}_{imp}}(\mathbf{k},\mathbf{k}+\mathbf{Q};\omega) G(\mathbf{k}+\mathbf{Q},\omega)$$

Signal at a given wavevector $\mathbf{Q}$ in the QPI can therefore arise either from the Green's function (bandstructure) of the material or from structure in the T-matrix.

We first consider the effect of the experimentally determined NaFeAs bandstructure on QPI. For the case of the structureless T-matrix, the structure in the QPI pattern is found to be well captured by the joint density of states (JDOS) $\int d\mathbf{k}\, \text{Im}[G(\mathbf{k},\omega)]\,\text{Im}[G(\mathbf{k}+\mathbf{Q},\omega)]$ [45]. The band structure of high temperature tetragonal NaFeAs observed in ARPES at low energies (<200 meV) [41,42] consists of nearly circular hole pockets at the Γ point and elliptical electron pockets at the X and Y points of the unfolded single Fe unit cell (see Supplementary Fig. S5). At low temperatures, long ranged nematic and spin density wave order sets in. As a result, the zone-center circular pockets become elliptical, the two zone-face pockets become inequivalent, and the unit cell doubles along the antiferromagnetic direction (which we take here to be X). This doubling produces band folding along the X direction, mapping the electron pockets at X onto the hole pockets at Γ so that a spin density wave gap appears in locations where the folded bands cross. This folding process leads to a Fermi surface characterized by a central elliptical region from the unhybridized hole pocket and four pockets arising from the SDW Fermi surface reconstruction. The outer pockets along Γ-X are located at $\mathbf{q}_\mathbf{D}$. They are protected by orbital symmetry from SDW gap opening and form Dirac cones [46]. This low temperature Fermi surface is shown in Fig. 4d, modeled from fits to ARPES data [41].

The calculated JDOS created by autocorrelation of the SDW-phase ARPES data is shown in Fig. 4e. This is an excellent match to the STS data at low temperatures as shown in Fig. 4f, the top half of which is the STS data from Fig. 4a and the bottom half of which is the ARPES JDOS from Fig. 4e. Both the qualitative shape of the STS image and the magnitudes of the scattering wavevectors seen in the STS data are captured by the SDW-phase ARPES JDOS. The origins of the scattering wavevectors seen in the STS data in Fig. 4a are simple to understand based on the ARPES Fermi surface. The four peaks at angles of ~45° to the Fe-Fe bonds (magenta dotted circle, Fig. 4f) are located at the wavevectors that correspond to scattering between the pockets at $\mathbf{q}_\mathbf{D}$ and the outer pockets along Γ-Y (magenta arrows in Fig. 4d). The intensity enclosed by the yellow dotted line in Fig. 4f along the Γ-Y axis is produced by scattering amongst the pockets aligned along the same axis (yellow arrow in Fig. 4d). The only feature that is not observed is scattering between the outer pockets at $2\mathbf{q}_\mathbf{D}$, indicating a suppression of scattering between Dirac cones. The observed STS features are thus well described by the Fermi surface reconstruction due to (π,0) order. We note that while spin density wave order is a natural explanation for this reconstruction, other orders with the same wavevector (such as antiferroic orbital ordering) can give rise to the same reconstruction of the band structure.

The issue we need to address is why these features persist up to 90K, above $T_S$ and $T_{SDW}$, where the system is globally tetragonal and paramagnetic. The Fermi surface observed at these temperatures by ARPES is $C_4$ symmetric, giving rise to a tetragonal symmetric JDOS, at variance with our STS data of Fig. 4c. At intermediate temperatures, between $T_{SDW}$ and $T_S$, SDW order is absent and the band folding does not occur. Yet, due to Q=0 ferro-orbital order, the Fermi surface is only $C_2$ symmetric, as observed by ARPES. In Fig. 4g, we show the JDOS arising from this orthorhombic paramagnetic Fermi surface, as determined by ARPES[41]. Although the JDOS still shows $C_2$ instead of $C_4$ symmetry, the intensity drops off monotonically away from $\mathbf{Q_x} = 0$, unlike what is observed in experiment (Fig. 4b). To illustrate this clearly, we plot in Fig. 4h line cuts through the STS data at $\mathbf{Q_y} = 0.1\dfrac{\pi}{a_0}$ (orange dotted line in Fig. 4a), which clearly show three peaks at all temperatures. To compare, we plot in Fig. 4i equivalent line cuts of the joint density of states from ARPES data obtained in the SDW, orthorhombic paramagnetic (OPM), and tetragonal paramagnetic (TPM) state, respectively. We see that the key feature required to produce the three peaks in the line cuts is the presence of the Fermi surface reconstruction with the wavevector $\mathbf{Q} = (\frac{\pi}{a_0}, 0)$, which produces disconnected regions of the Fermi surface separated by $\mathbf{q_D}$.

Therefore, the observed sample-averaged band structure alone cannot account for the observed STS features at high temperatures. From Eq. (1), two scenarios may be put forward to explain the data: additional structure in the T-matrix, for example due to the impurity nucleating a droplet of ordered phase in its vicinity, or self-energy corrections to the Green's function due to strong local stripe fluctuations. To gain insight into these two alternatives, we use a simplified band model with two parabolic bands forming a hole and an electron Fermi pocket. While this model is too simple to explain all of the details of the experimental data, we will show that it accounts for the crucial qualitative effect, namely the persistence to high temperatures of intensity peaks at $\pm\mathbf{q_D}$.

We begin with low temperatures. The reconstructed Fermi surface in the SDW phase is shown in Fig. 5a, and the resulting QPI is shown in Fig. 5b. Comparison of Fig. 4d and Fig. 5a shows that the key difference between the model and the data is the absence, in the model, of the unhybridized elliptical hole pocket centered at the Gamma point. This pocket produces the central rod structure seen in Fig. 4e which is absent in Fig 5b. However the peaks at ±45 and ±135 degrees arising from the hybridization of the electron and hole pockets are present in the simplified model and in the data.

We now move to the paramagnetic state. We find (calculation details in Supplementary Information) that calculations based on combining the noninteracting Green's functions of the high temperature phase (either tetragonal or orthorhombic) with a structured T-matrix corresponding to a local region of SDW order nucleated by the impurity bear no resemblance to the long range SDW QPI calculation in Fig. 5b . We therefore consider here a model in which the T matrix is structureless but the electron Green's function includes a self-energy correction due to *unidirectional*, short-ranged, SDW fluctuations. In this case, we parameterize the short-ranged correlations by the SDW correlation length ξ and use for the Green's function an ansatz originally introduced by Lee, Rice and Anderson to describe short ranged order in charge density wave materials [47]. By increasing ξ to modest values, we find that the calculated QPI spectrum acquires the main features displayed by the QPI in the SDW phase, even though no long-range order is present. For instance, in Fig. 5c we present the QPI for ξ = 8$a_0$, where we see similar features as those in the calculated QPI of the SDW phase. This becomes more transparent by analyzing line cuts along the orange line of Fig. 5c. We see in Fig. 5d that as ξ increases, the peak at $\mathbf{Q_x} = 0$ splits, moving to the same positions as the QPI peaks in the SDW phase. This is in qualitative agreement with the temperature evolution of the experimental QPI cuts shown in Fig. 4i.

Further insight into the nature of the electronic interactions in NaFeAs is obtained by quantitatively measuring the $C_4$ symmetry breaking as a function of energy and temperature. In order to quantify the degree of $C_4$ symmetry breaking in a given LDOS image, we first crop a small region around each defect in a LDOS image and average all regions together to produce the average spectroscopic image around a single defect. We then rotate the averaged image by $90^0$, subtract it from the original averaged image, and normalize by the sum of the rotated and unrotated images. This procedure is depicted in Fig. 6a where we show the unrotated and rotated 26 K defect images at E=10 meV and the resulting "anisotropy map." If the original image is $C_4$ symmetric, then the anisotropy map is zero at every pixel. As such, any non-zero value in the map corresponds to $C_4$ symmetry breaking. Larger positive or negative values in the anisotropy map correspond to more intense anisotropy. Shown in Fig. 6b-d are three such anisotropy maps from the same temperature as Fig.6a at three different energies. We can clearly see the degree of $C_4$ symmetry breaking decreases with increasing energy. To describe this effect as a function of temperature and energy using a single number (which we call the anisotropy parameter), we sum the absolute value of the intensity in the anisotropy map which we plot in Fig. 6e for several energies at five different temperatures. We note that the anisotropy is strong close to the Fermi energy, peaks slightly above (E=10 meV), and decreases in strength with increasing temperature. A direct comparison of the energy dependence of the anisotropy parameter with the average spectrum in the SDW state (shown in the inset to Fig. 6e) shows that they share a common energy scale. We take this as another indication that both phenomena share a common origin in the spin physics of the pnictides.

The breaking of local $C_4$ symmetry above $T_S$ in the low-energy electronic states, and the association of these features with unidirectional fluctuations (i.e. only one of $(0, \pi)$ or $(\pi, 0)$) are the key findings of this experiment. We suspect that our samples, which are not detwinned, have small, residual strain fields which nucleate the large amplitude $C_4$ symmetry breaking effect reported here. The more far-reaching implications come from the analysis of the QPI patterns, which clearly indicates that large amplitude fluctuations with a wavevector $\mathbf{Q}_a = (\frac{\pi}{a_0}, 0)$ are responsible for the anisotropic electronic structure locally observed in the high temperature state while fluctuations at the orthogonal wave vector $\mathbf{Q}_b = (0, \frac{\pi}{b_0})$ are suppressed. The selection of one fluctuation channel over another is generally believed to be a nonlinear effect signaling that spin fluctuations in one direction have large enough amplitude to suppress fluctuations in the other[16]. While both antiferroic orbital and antiferromagnetic fluctuations can produce these fluctuations, our finding that the energy dependence of the electronic anisotropy matches the low temperature spin density wave gap makes spin fluctuations the natural explanation for the observed anisotropy. Such a redistribution of electronic spectral weight by short-range magnetic order at high temperature is consistent with neutron scattering [48] and nuclear magnetic resonance [49] measurements that observe a clear increase of magnetic fluctuations around the same temperature scale of 90 K in NaFeAs. It is also consistent with the absence of Fermi surface reconstruction in this temperature regime as observed by ARPES, since the zero-energy states are not gapped by short-range SDW order. Interestingly, recent muon spin rotation (μSR) data [50] on another family of iron pnictides found evidence of such short-range SDW order even above the magnetic and structural transitions. The presence of these fluctuations, even far from any phase boundary, must be taken into account in theories of the anomalous pnictide normal state and of the superconductivity arising from it.

**Acknowledgements:**

We thank L. Zhao and C. Gutierrez for experimental help and T. Giamarchi and I. Eremin for discussions. We thank M. Yi and Z-X. Shen for sharing ARPES data on NaFeAs. This work is supported by the National Science Foundation through the Partnerships for International Research and Education grant no OISE-0968226. Equipment support is provided by the Air Force Office for Scientific Research under grant no. FA9550-11-1-0010. Support is also provided by the Department of Energy Basic Energy Sciences program through grant DOE-ER-046169 (AJM), Defense Advanced Research Projects Agency grant no. N66001-12-1-4216 (ANP, ER) and the National Science Foundation of China and Ministry of Science and Technology of China (LYX, XCW, CQJ).

The authors declare no competing financial interests.

Correspondence and requests for materials should be addressed to ANP (apn2108@columbia.edu)

**Figure Captions:**

**Figure 1 | Topographic STM and STS maps of LiFeAs and NaFeAs at low temperatures. a** Constant current STM topograph of LiFeAs (V=-120 mV, I =270 pA, T=39 K). The alkali atoms are observed on the surface layer, and the underlying iron atoms are illustrated in the lower left side. The two inequivalent positions of the iron atoms are shown with filled and unfilled circles. An alkali vacancy is identified with a black arrow, and red lines indicate the size and orientation of iron site defects. Gray arrows identify the Alkali-Alkali directions, and cyan arrows identify the orientation of the Fe-Fe lattice (continued for rest of Fig. 1). **b**, Differential conductance (dI/dV) map (V=-120 mV, I = 270 pA, T=39 K) of LiFeAs. The iron defects have prominent signatures in the dI/dV maps with the same size and symmetry as the topographic features, as indicated by the red lines. **c**, dI/dV map of NaFeAs (V=-100 mV, I=300 pA, T=26 K) showing prominent features that are oriented along one Fe-Fe direction which is at 45° to the crystallographic axes. Size and orientation of iron site defect that produce the features are indicated by red lines. **d-i**, dI/dV maps at different energies of the area shown in **c**, showing a clear variation in the contrast and size of features as a function of energy. **j** Large area dI/dV image of NaFeAs (same junction conditions and temperature as **d**) showing universality of spectroscopic features and the presence of domains where the unidirectional features rotate by 90°.

**Figure 2 | STS maps of NaFeAs in real and Fourier space in the SDW phase. a-d**, Large-area differential conductance maps on NaFeAs (V=-100 mV, I=300 pA, T=26 K). Arrows in **c** indicate ferromagnetic (FM)

and antiferromagnetic (AFM) directions. **e-h** Corresponding FFT images. The FFT images show well-defined structure whose wavelengths and intensities are energy dependent. Size of images is half of the single Fe unit cell Brillouin zone.

**Figure 3 | Temperature dependence of anisotropic STS features in real and Fourier space. a,c,e,g,i,k**, Large area maps of the differential conductance at 10 meV at different temperatures on NaFeAs. The raw images show that the unidirectional features persist up to the highest temperatures shown. However, the intensity of the unidirectional features decreases with increasing temperature and becomes weak above 80 K. **b,d,f,h,j,l**, The corresponding FFT images (same scale as Fig. 2e-h). Junction settings for 26 K, 38 K, 46 K, and 61 K are V=-100 mV and I=300 pA. Junction settings for 54 K and 75 K are V=-50 mV and I=300 pA. It is seen that the same basic structure exists in all the FFT images even above $T_{SDW}$=39 K and $T_S$=54 K.

**Figure 4 | Comparison between STS and ARPES joint density of states. a-c**, Fourier transforms of STS images at 26, 46 and 61 K respectively. All three STS images show the presence of strong scattering intensity at $\mathbf{q_x} = 0$ and $\mathbf{q_x} = \pm \mathbf{q_D}$ (denoted by green dot). **d**, ARPES intensity at the Fermi surface in the SDW (T<$T_{SDW}$) phase (from ref. 38). Magenta lines indicate scattering that produces 45° peaks in JDOS, and yellow arrow corresponds to 0/180° intensity. **e**, Joint density of states (JDOS) from autocorrelation of SDW ARPES intensity. **f**, SDW STS from **a** placed over SDW JDOS from **e**. Color-coded dotted-line contours correspond to scattering peaks that are produced by same colored scattering vectors in **d**. **g**, JDOS for OPM phase. Fermi surface shown in Supplementary Fig. S5. **h,i**, Line cuts of JDOS and STS along dotted orange lines in a and e for different phases/temperatures. Clear peaks arising from reconstructed Fermi surface scattering in STS data only matches SDW JDOS.

**Figure 5 | Theoretical bandstructure and short-ranged SDW calculations. a,b**, Model Fermi surface density of states ($n(\mathbf{k})$) in **a**, and QPI ($|\delta n(\mathbf{k})|$) in **b** in the presence of long-range SDW order. **c**, QPI in the presence of short-range (ξ = 8$a_0$) SDW order (see Supplementary Information ). **d**, Line cuts along orange dotted line in **c** for different values of correlation length (in units of $a_0$). Split peaks develop with increasing correlation length in analogy to STS data.

**Figure 6 | Energy and temperature dependence of electronic anisotropy. a**, Visual depiction of anisotropy map calculation procedure. Average defect image is rotated by 90° and subtracted from the unrotated image and normalized by the sum of the rotated and unrotated images. This procedure results in non-zero values in space where $C_4$ symmetry is broken. Image was measured in the SDW phase at 26 K. **b-d**, Anisotropy maps at various energies calculated at the same temperature as **a** and plotted on the same colorscale. The strength of the anisotropy is seen to decrease with increasing energy. **e**, Total anisotropy as a function of energy at various temperatures (junction conditions the same for all temperatures). Each data point is generated by summing the absolute value of the appropriate anisotropy map such as those shown in **a-d**. It is seen that the anisotropy is maximum at an energy E=10 meV and falls off at higher energy. The average spectrum of the sample at each temperature is shown in the inset. The maximum of the anisotropy is located close to the midpoint of the gap, and the energy range of the anisotropy is comparable to the size of the low-temperature gap.

# Figure 1

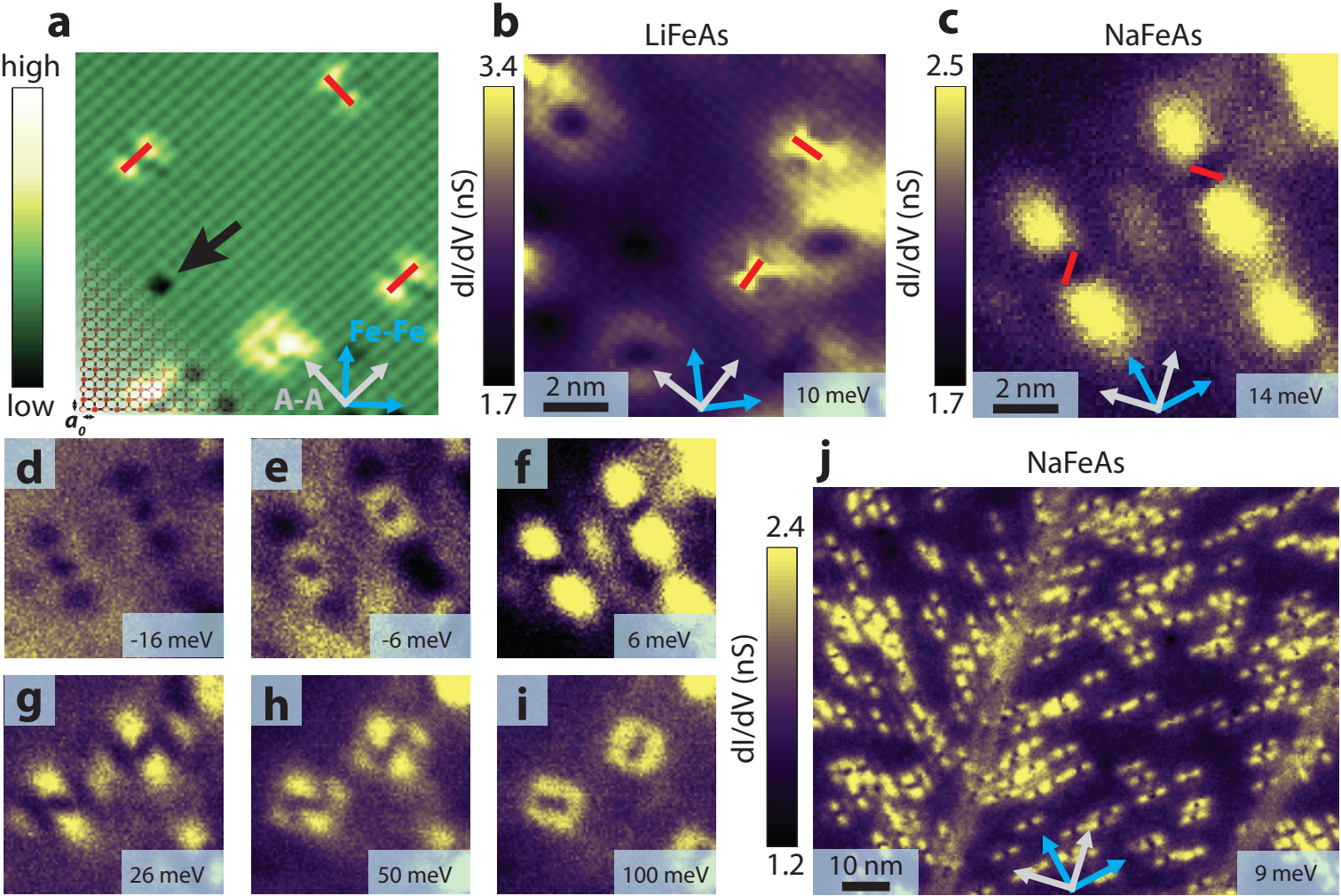

# Figure 2

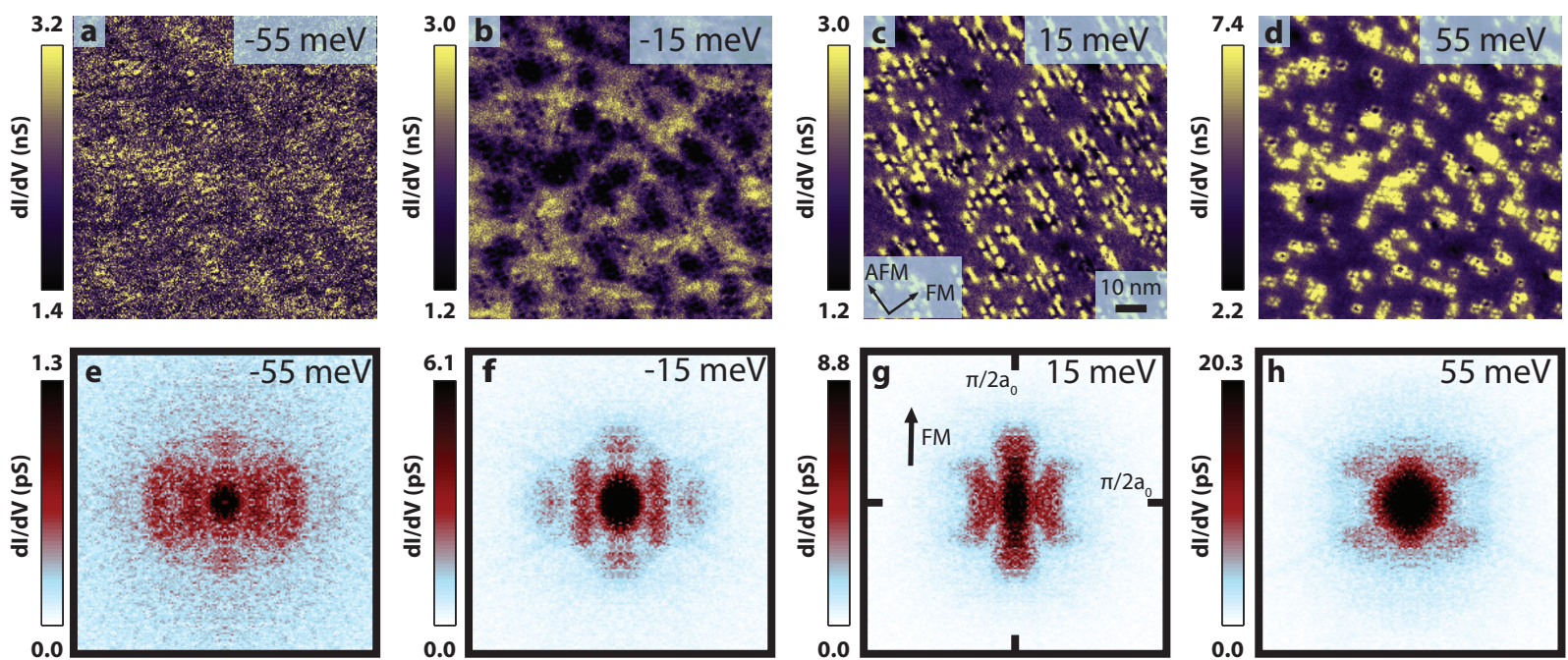

# Figure 3

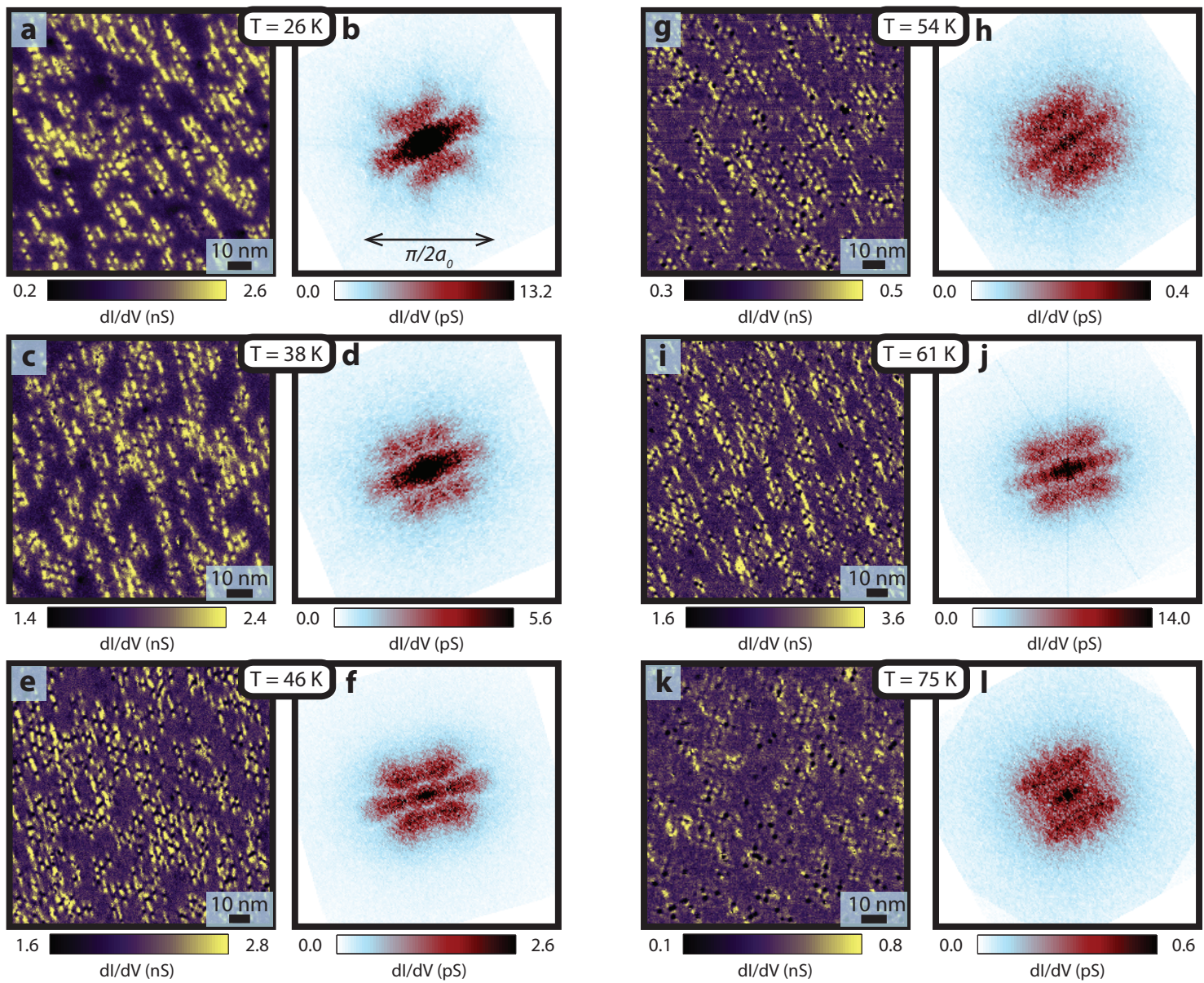

Figure 4

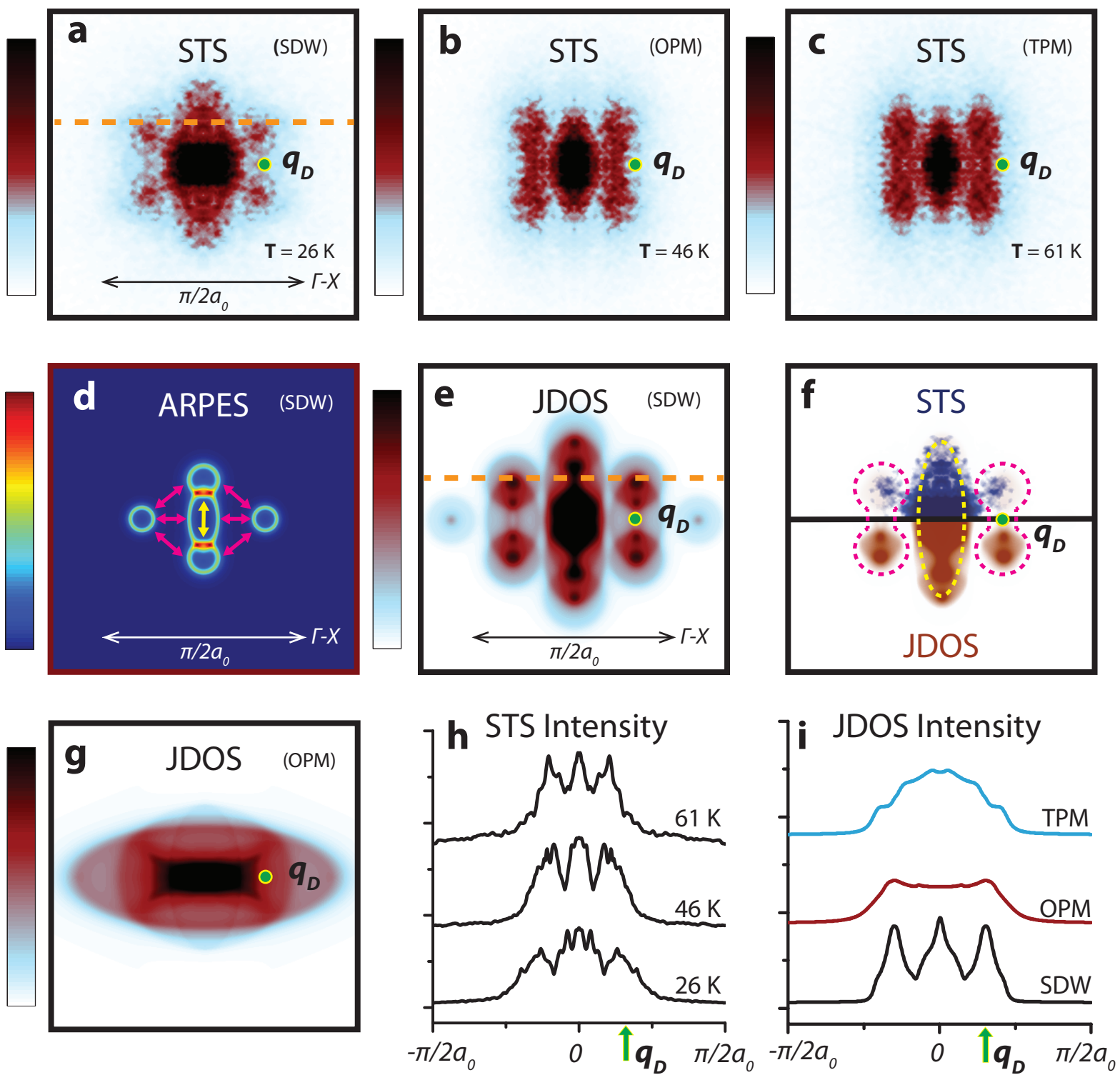

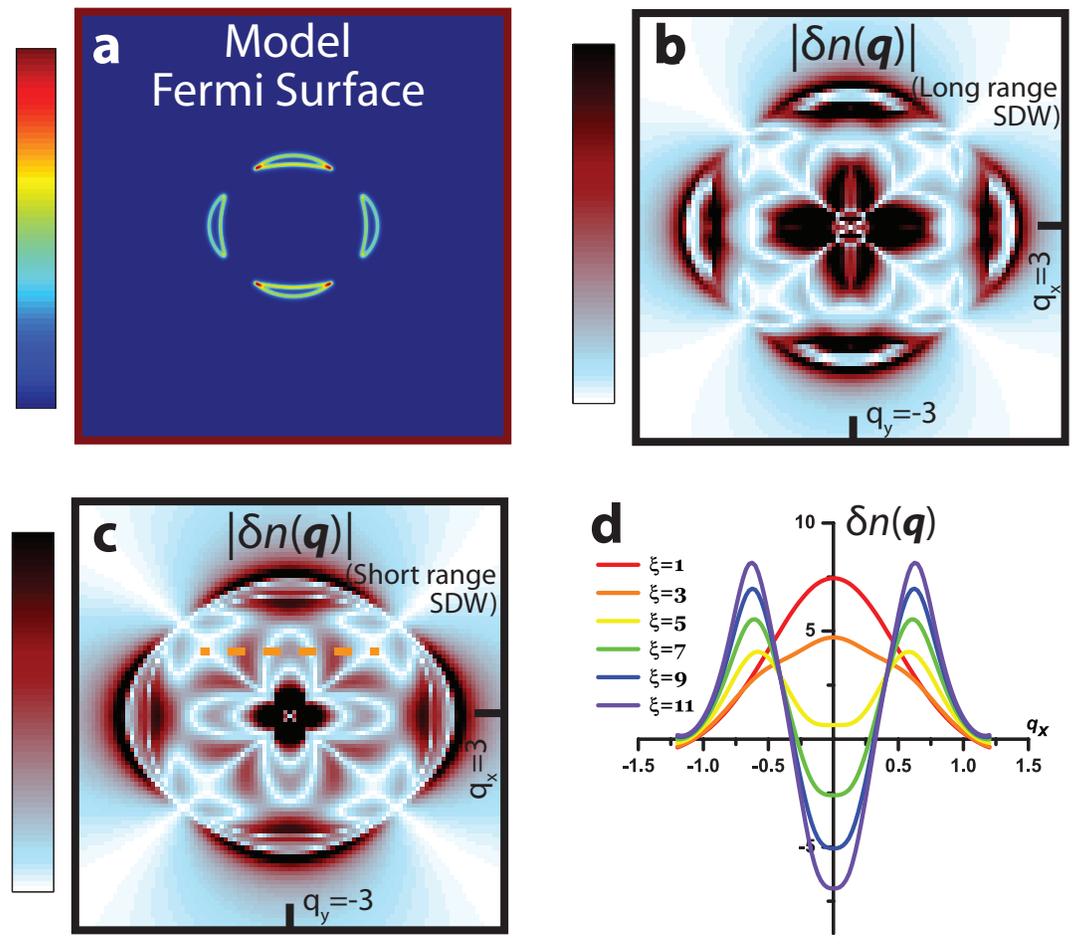

# Figure 6

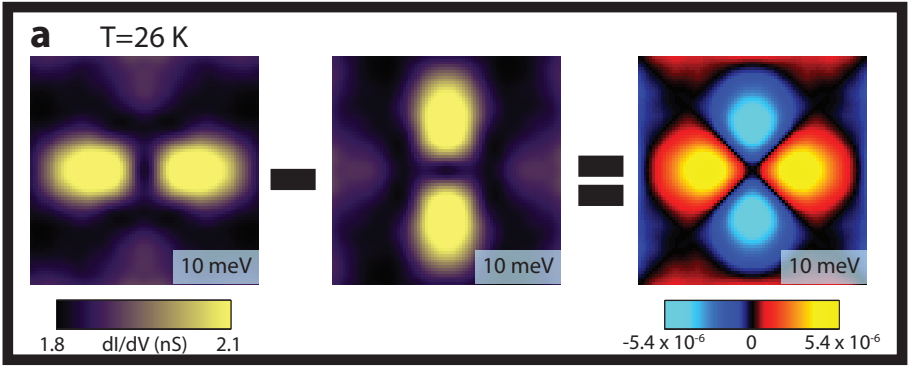
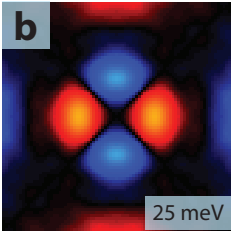
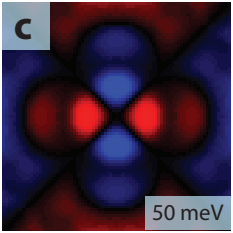
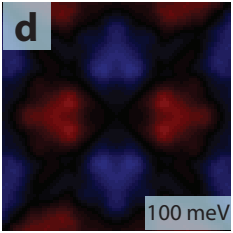
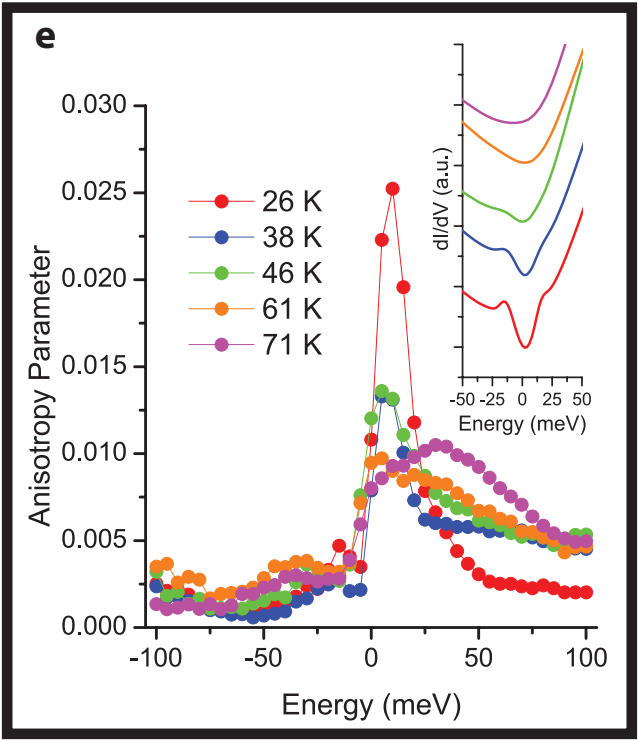

# Supplementary Information for

# Visualization of Electron Nematicity and Unidirectional Antiferroic Fluctuations at High Temperatures in NaFeAs

E. P. Rosenthal[1], E. F. Andrade[1], C. J. Arguello[1], R. M. Fernandes[2], L. Y. Xing[3], X. C. Wang[3], C. Q. Jin[3], A. J. Millis[1], A. N. Pasupathy[1]

[1]Department of Physics, Columbia University, New York NY 10027, USA

[2]School of Physics and Astronomy, University of Minnesota, Minneapolis, MN 55455, USA

[3]Beijing National Laboratory for Condensed Matter Physics, Institute of Physics Chinese Academy of Sciences, Beijing 100190, China

## Materials and Methods

The single crystals of LiFeAs and NaFeAs used for this study were grown by self flux methods. $Li_3As$ and $Na_3As$ precursors were synthesized by the reaction of Li or Na lumps with As powder at $600°C$ for 10 hours in an evacuated quartz tube. The FeAs precursor was prepared by sintering the mixture of Fe and As powders at $750°C$ for 20 hours in an evacuated quartz tube. Later, the $Li_3As$ or $Na_3As$, FeAs and As powders were mixed according to the stoichiometric ratio of $LiFe_{0.3}As$ or $NaFe_{0.3}As$. The powder mixture was pressed into a pellet in an alumina oxide tube and sealed in a Nb tube with Argon gas at a pressure of 1 atm before being sealed in an evacuated quartz tube. This quartz tube was then heated up to a temperature to $1100°C$ for the $LiFe_{0.3}As$ compound ($950°C$ for $NaFe_{0.3}As$) for 10 h and then cooled down to $700°C$ at a rate of $5°C$ per hour. We obtained LiFeAs and NaFeAs crystals up to 5 mm × 5 mm × 0.5 mm in size. All the preparation work was carried out inside a glove box protected from air with high purity Ar gas.

STM and STS measurements were carried out using a variable temperature, ultra high vacuum, home-built STM with millikelvin temperature stability and sub-picometer position accuracy. Single crystal samples were prepared for STM measurements in an Ar glove box, loaded into the microscope, and then cleaved *in situ* between temperatures of 20 K and 80 K. Conductance maps were obtained using standard lock-in amplifier techniques with an oscillation amplitude of 1.5mV and a lock-in frequency of 1.666 kHz.

## Supporting Online Text

### I. Comparisons Between Topographic and Spectroscopic Images in LiFeAs and NaFeAs

Figure S1 displays the topographic image associated with fig. 1b along with STS images at various energies. The $C_4$ symmetric alkali vacancies maintain their full symmetry over the range of energies measured. Similarly, the iron defects maintain their respective $C_2$ symmetry aligned with the Li-Li bonds over the range of energies measured. The radial extent of the spectroscopic features associated with each defect is limited to a few lattice constants. Figure S2 shows the topographic image associated with fig. 1c, which has been reproduced in fig. S2b. The iron site defects are identified with red lines, and the spectroscopic features in fig. S2b are seen to originate from these defects, though their size and orientation are uniformly different.

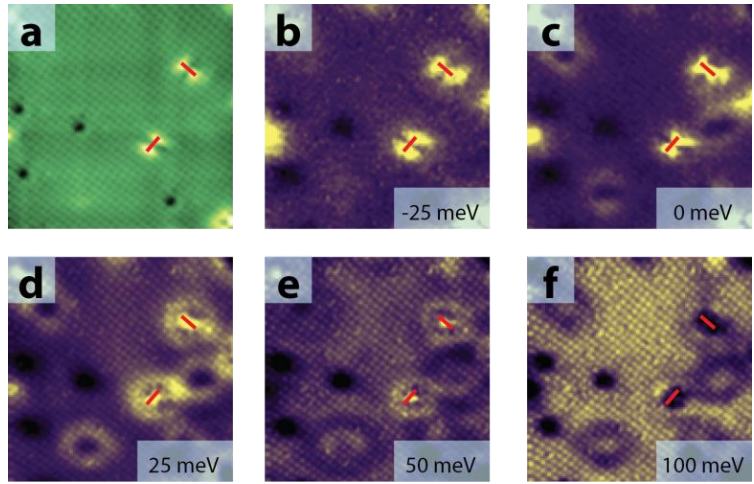

**Fig. S1: a**, Topographic image of LiFeAs corresponding to the same area as Fig. 1b. Iron site defects are identified by red lines as in the main text. **b-f**, Conductance maps corresponding to the same are as **a** and Fig. 1b at various energies. All experimental conditions are the same as Fig. 1b.

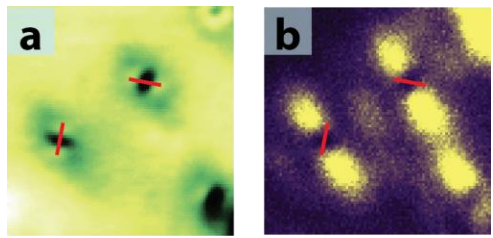

**Fig. S2: a**, Topographic image of NaFeAs corresponding to same area as Fig. 1c and reproduced here in **b**. Iron site defects are identified with red lines as in the main text.

## II. Association of Defects with Spectroscopic Features

Intrinsic defects in NaFeAs are visible in topographic images, as shown in Fig. S3a. By marking each defect from the topography, one may compare the locations of the defects with the locations of unidirectional features seen in simultaneously obtained spectroscopic maps (Fig. S3b, c). There is a clear correlation between the locations of the defects and the unidirectional features, with the unidirectional features appearing centered about the intrinsic defects. Importantly, the spectroscopic feature seen in dI/dV maps is not dependent on the nature of the defect seen in the topography, indicating that the features seen in dI/dV maps are not specific to a particular kind of defect.

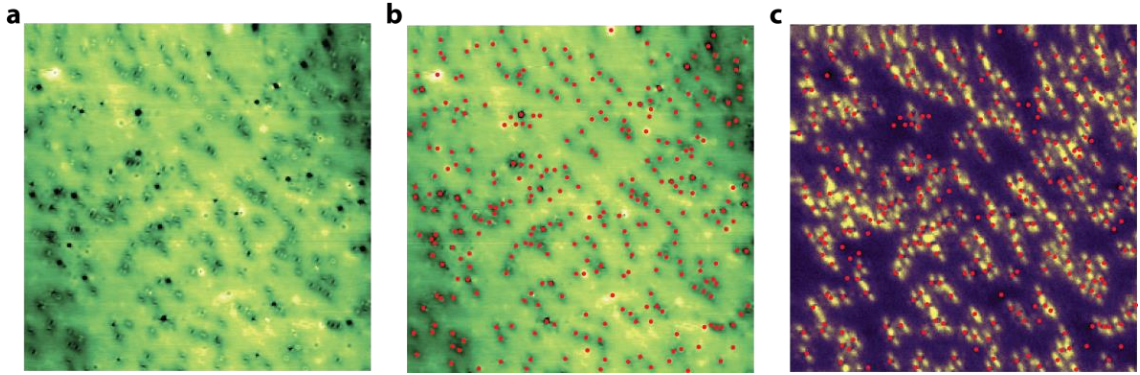

**Fig. S3:** Comparison between defect locations and spectroscopic, unidirectional features. **a**, Topography at T=26 K showing variety of intrinsic defects. **b**, Same image as **a** but with red dots identifying the location of each defect. **c**, Simultaneously obtained conductance map at E=10 meV with defect locations overlain.

### III. Processing of FFT Images

Fourier transform images were processed in order to increase the signal to noise of QPI features. First, drift was removed from STM spectroscopic maps by affine transformation, using the simultaneously acquired topographic image. After the affine transformation, the conductance images were Fourier transformed. The Fourier image was then mirror symmetrized along the X and Y lattice directions.

### IV. Visualization of ARPES Data

The ARPES Fermi surface and joint density of states (JDOS) shown in Fig. 4 are fits to raw data from ref. 1 on NaFeAs along the Γ-X and Γ-Y directions in all three phases – tetragonal paramagnetic (TPM), orthorhombic paramagnetic (OPM), and spin density wave (SDW). A comparison of the published ARPES data and the fitted model in the SDW state are shown in Fig. S4 a and b, respectively. The Fermi surface and corresponding JDOS for the TPM phase, as well as the Fermi surface used for the OPM JDOS in Fig. 4g, is shown in Fig. S5.

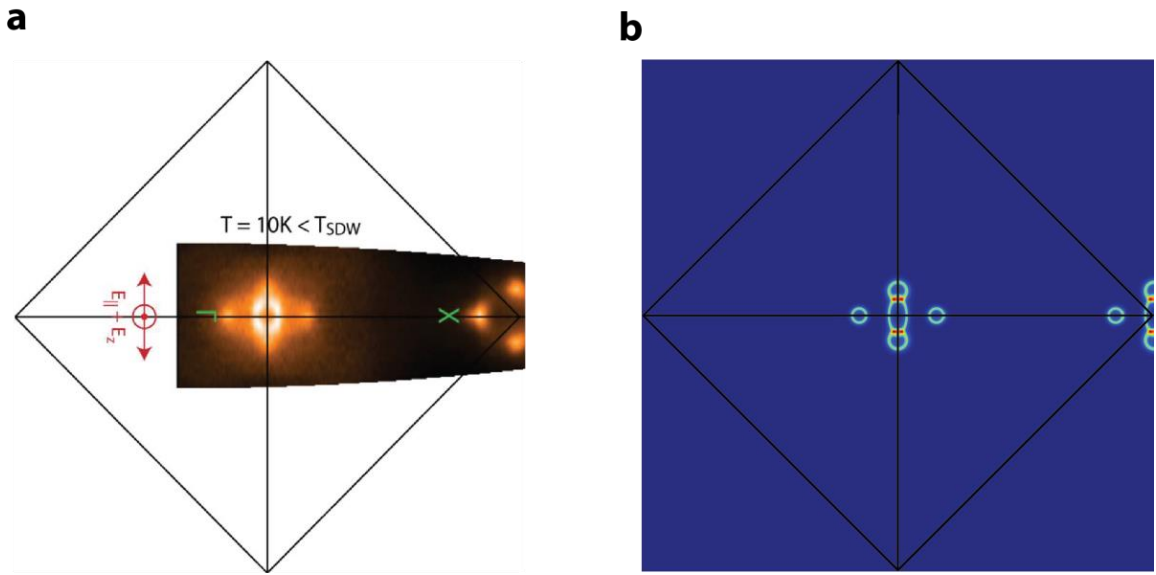

**Fig. S4:** Comparison between ARPES data and model. **a**, ARPES Fermi surface from ref. 1. Black box corresponds to two Fe Brillouin Zone. **b**, Model Fermi surface generated from curves fit to **a**.

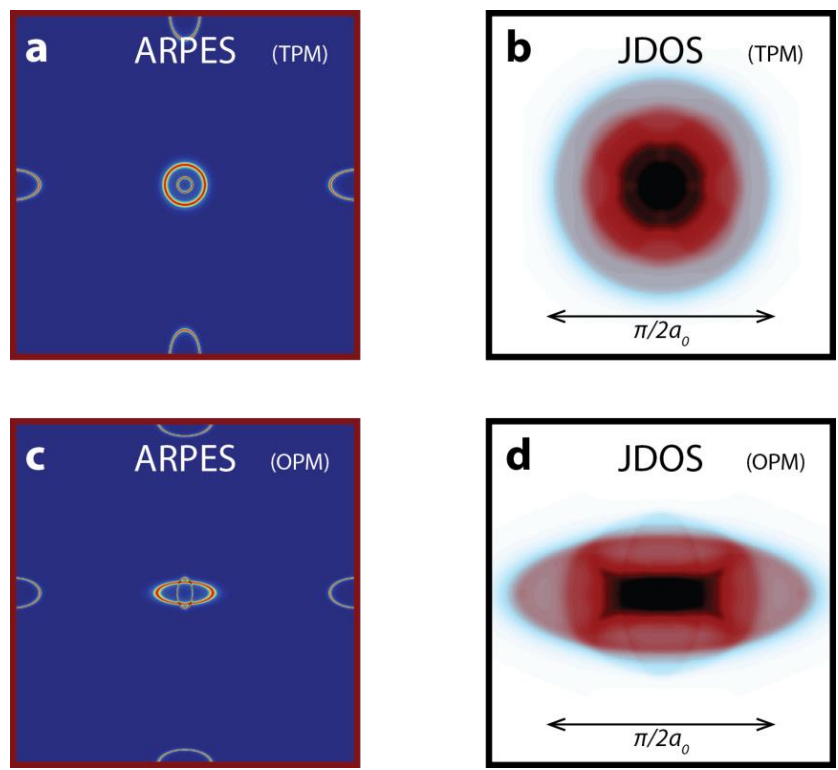

**Fig. S5:** ARPES Fermi surface and JDOS for tetragonal paramagnetic (TPM) and orthorhombic paramagnetic (OPM) phases. **a,b**, TPM ARPES Fermi surface from reference ref. 1. and corresponding JDOS calculated by autocorrelation. **c,d**, OPM Fermi surface and JDOS.

## V. Calculation of $|\delta n(q)|$ for Figure 5

In order to calculate the theoretical quasiparticle interference (QPI) in Fig. 5b, we used Eq. 1 with the T-matrix restricted to the leading and quadratic order in the impurity potential $V$:

$$T(\mathbf{k}, \mathbf{k}+\mathbf{q}; \omega) = V(\mathbf{q}) + \int d\mathbf{p}\, V(\mathbf{p}) G(\mathbf{k}+\mathbf{p}, \omega) V(\mathbf{q}-\mathbf{p}) + \cdots$$

For simplicity, we consider a two-band model with a circular hole pocket and an elliptical electron pocket separated by the SDW ordering vector $\mathbf{Q}$:

$$\varepsilon_{1,\mathbf{k}} = \epsilon_1 - \frac{k^2}{2m}$$

$$\varepsilon_{2,\mathbf{k}+\mathbf{Q}} = -\epsilon_2 + \frac{k_x^2}{2m_x} + \frac{k_y^2}{2m_y}$$

such that $G_a^{-1} = \omega - \varepsilon_{a,\mathbf{k}} + i0^+$ in the paramagnetic tetragonal state. We used the parameters $\epsilon_1 = \epsilon_2$, $m_x = 2m_y$, and $\sqrt{m_x m_y} = 1.01m$. In the SDW state, the normal part of the Green's function is given by:

$$G_a = \frac{\omega - \varepsilon_{\bar{a}}}{(\omega - E_1)(\omega - E_2)}$$

where:

$$E_{1,2} = \left(\frac{\varepsilon_1 + \varepsilon_2}{2}\right) \pm \sqrt{\left(\frac{\varepsilon_1 - \varepsilon_2}{2}\right)^2 + M^2}$$

are the quasi-particle excitation energies and $M$ is the magnetic order parameter set to $M = 0.1\sqrt{\epsilon_1 \epsilon_2}$. The resulting Fermi surfaces for both the normal and SDW state are shown below in Fig. S5.

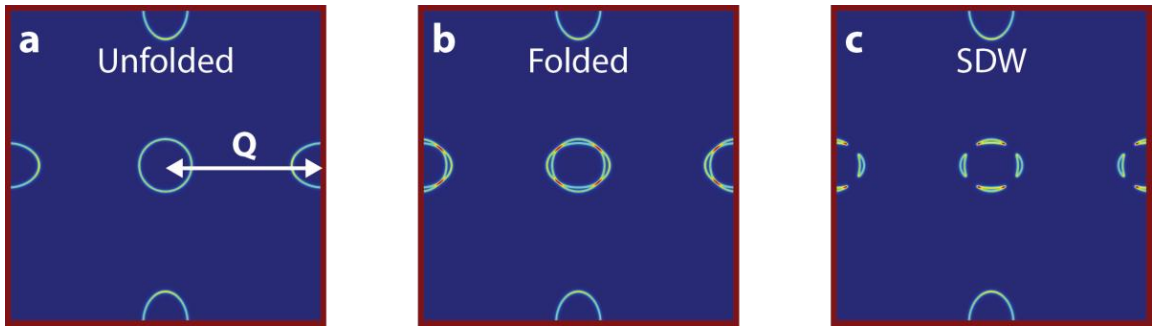

**Fig. S6:** Fermi surfaces for two-band model. **a**, First Brillouin zone for single Fe unit cell. A circular hole pocket is at the center, and elliptical electron pockets are at the Brillouin zone edge which is located a distance $\mathbf{Q}$ from the zone center. **b**, Brillouin zone folded by SDW ordering vector $\mathbf{Q}$. **(C)** First Brillouin zone in the presence of long-range SDW order. Gaps are opened in regions nested by the SDW ordering vector.

As was mentioned in the text, we consider two possible scenarios of magnetic correlations in the paramagnetic phase: a magnetic droplet and short-range SDW order. In the former case, the impurity potential $V(\boldsymbol{q})$ is peaked at the SDW ordering vector $\pm\boldsymbol{Q}$:

$$V_\sigma(\boldsymbol{q}) = \frac{1}{2}\left(\frac{\lambda}{|\boldsymbol{q}-\boldsymbol{Q}|^2+\lambda^2} + \frac{\lambda}{|\boldsymbol{q}+\boldsymbol{Q}|^2+\lambda^2}\right)\sigma$$

Here, $\sigma$ is a spin index, indicating that the droplet potential has a directional dependence. Because of this, summing over spin indices cancels the linear term in Eq. (S1), and we need to consider the term quadratic in $V$. In our calculations, we used $\lambda = 0.2 k_F$.

In the case of short-range magnetic order in the paramagnetic phase, we consider the Lee-Rice-Anderson (*2*) Green's function:

$$\tilde{G}_a^{-1} = G_a^{-1} - \frac{M^2}{\omega - \varepsilon_{\bar{a}} + i\xi^{-1}}$$

where $\xi$ is proportional to the magnetic correlation length. We fix M to have the same value as in the SDW phase. We consider two particular situations: one with a finite $\xi$ and the other in the limit $\xi \to \infty$. The latter is equivalent to the case of long-range SDW order *without* the coherence factors that mix the hole and electron states. Except in the magnetic droplet case, the impurity potential is assumed to be that of a point impurity, i.e. independent of $\boldsymbol{q}$.

Within this model, we obtain Fig. 5b for the $\xi = 8$ short-range SDW QPI. This image is repeated in Fig. S7 where we include the other three scenarios: $\xi \to \infty$, long-range SDW order, and magnetic droplet. The corresponding real space density of states (DOS) has been plotted next to each respective scenario. Not only is it clear that the short-range SDW order in the paramagnetic phase can produce similar QPI to that of the long-range order, but the unidirectional spectroscopic features seen in real space are qualitatively similar to the real space calculated QPI. Conversely, the magnetic droplet model bears no resemblance to either the calculated long-range SDW QPI or the STS data.

One of main features of the experimental QPI spectrum is the existence of maxima located at angles of ~45° to the Fe-Fe bonds (Fig 4a), which persist up to high temperatures. They can be visualized in a more transparent way via the constant-$k_y$ cut shown in Fig. 4h (correspondent to the dotted orange line in Fig. 4a), which displays maxima at $k_x=0$ and $k_x=q_{D,x}$. Our calculated QPI also displays the same maxima in the SDW phase (see Fig. S6e). Our results reveal that, in the paramagnetic phase, these maxima only appear if the magnetic correlation length overcomes a certain threshold – otherwise, there is only the maximum at $k_x=0$. This is shown in Fig. 5d, where the theoretical constant-$k_y$ cut correspondent to the dotted orange line in Fig. 5c is plotted for several different values of $\xi$. As the magnetic correlation increases, the maximum is displaced from $k_x=0$ to $k_x=q_{D,x}$. For the case of a magnetic droplet, the maximum remains at $k_x=0$. Thus, the experimental QPI is consistent with the presence of strong unidirectional magnetic fluctuations.

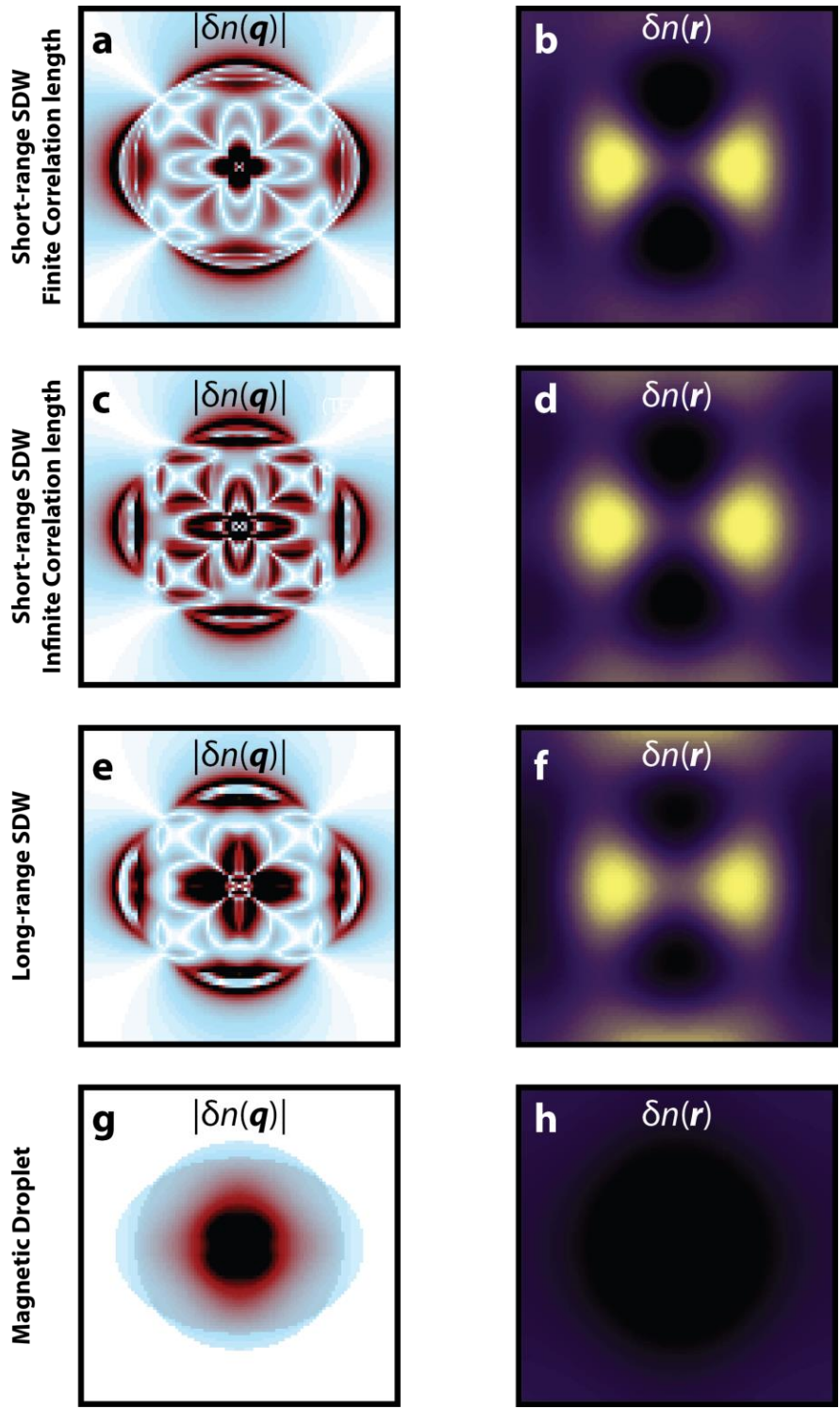

**Fig. S7:** Models of QPI with magnetic order. Fourier space DOS is on the left, and real space DOS is on the right. **a,b,** Short-range SDW order using Lee-Rice-Anderson Green's function for correlation length $\xi = 8$. **c,d,** Same model as **a** and **b** but $\xi \to \infty$. **e,f,** Long-range SDW order. Real space DOS shows same unidirectional quality as short-range SDW calculations and STS measurements. **g,h,** Magnetic droplet calculation.

### V. Generation of Anisotropy Maps and Anisotropy Parameter

All conductance maps were first upconverted to four times the original spatial resolution. Defects were identified, and a circular area of diameter 6.7nm was cropped around each defect. All cropped defects were averaged together to create a supercell image of the average defect appearance for each energy and temperature. The supercells were then $C_2$ symmetrized. The anisotropy maps were calculated by taking the difference of the supercell and the supercell rotated by 90°. The anisotropy maps were normalized by first summing the rotated and unrotated supercell and then summing over all values. The anisotropy parameter was calculated by taking the absolute value of the rotated and unrotated supercell difference, summing all values, and dividing same normalization used in the maps. A perfectly $C_4$ symmetric would not change under 90° rotation and would yield an anisotropy parameter of zero. Alternatively, a maximally $C_2$ symmetric image would yield an anisotropy parameter of one.